\documentclass[10pt,twocolumn,reqno, amsmath,amssymb, aps,prb,superscriptaddress]{revtex4-1}

\usepackage[pdftex]{graphicx}
\usepackage{float}
\usepackage{perpage}

\usepackage{color}
\usepackage{rotating}
\usepackage[breaklinks=true,colorlinks,citecolor=blue,linkcolor=blue,urlcolor=blue]{hyperref}
\usepackage{appendix}
\usepackage{graphicx}
\usepackage{natbib}
\usepackage{dcolumn}
\usepackage{bm}
\usepackage{hyperref}
\usepackage[mathlines]{lineno}
\usepackage[caption=false]{subfig}

\begin{document}

\title{Multistability of a Josephson parametric amplifier coupled to a mechanical resonator}

\author{Olga Shevchuk}
\affiliation{Kavli Institute of Nanoscience, Delft University of Technology, Lorentzweg 1, 2628 CJ Delft, The Netherlands}
\author{Rosario Fazio}
\affiliation{NEST, Scuola Normale Superiore and Istituto di Nanoscienze-CNR, I-56126 Pisa, Italy}
\author{Ya. M. Blanter}
\affiliation{Kavli Institute of Nanoscience, Delft University of Technology, Lorentzweg 1, 2628 CJ Delft, The Netherlands}

\begin{abstract}
We study the dynamics of Josephson Parametric Amplifier (JPA) coupled to a mechanical oscillator, as realised with
a dc Superconducting Quantum Interference Device (SQUID) with an embedded movable arm. We analyse this system
in the regime when the frequency of the mechanical oscillator is comparable in magnitude with the plasma oscillation
of the SQUID. When the nano-mechanical resonator is driven, it strongly affects the dynamics of the JPA.  We show
that this coupling can considerably modify the dynamics of JPA and induce its multistability rather than common bistability.
This analysis is relevant if one considers a JPA for detection of mechanical motion.

\end{abstract}
\maketitle

\section{Introduction}
Recently, there has been considerable interest in coupling mechanical resonators to optical and microwave radiation, as well as to
electric conduction~\cite{Cleland,Poot,Aspelmeyer}. Resulting devices, nanoelectromechanical and optomechanical systems,
combine excellent mechanical and electrical/optical properties, such as low dissipation both in the cavity and in the mechanical resonator.
These systems have already displayed a rich variety of interesting physical phenomena. At the same time they  found applications
{\em e.g.} as sensors and transducers. Most of the experiments so far have been carried out in the regime of classical mechanical
motion, though recently the quantum regime has been demonstrated as well~\cite{Oconnell,Teufel,Chan,Safavi,Palomaki}.

An important problem in this field is to find efficient schemes to detect the mechanical motion. To this end, one should find
systems whose properties are significantly affected by the mechanical resonator, carrying  consequently distinct signature of this
coupling. On the other side the coupling to the detector has  sizeable consequences on the resonator too. Backaction, understood
as an effect of the detector (for example, an optical or a microwave cavity) on the properties of the mechanical resonator is one of
the fundamental issues in the field of nano- and optomechanics~\cite{Poot}. Even though backaction is not always wanted in
the experiment and can be suppressed with backaction-evasion techniques~\cite{noisesqueezing}, one can nevertheless use it
as an advantage in order to manipulate, for example to cool~\cite{cooling} and to heat the resonator. In optomechanics backaction
is provided by radiation pressure and its properties are very well established, both theoretically and experimentally~\cite{Aspelmeyer}.
More recently, backaction in nanomechanical devices, caused by electrostatic interactions \cite{Meerwaldt,Koenig} and by Lorentz
force \cite{TunableBackaction,Etaki}, both in the classical regime, has been demonstrated experimentally. It is important
that in all these examples, the mechanical resonators were in the linear regime. The backaction effect consisted in the
modification of the frequency (optical spring) and the quality factor, as well as in the induced non-linearity of the mechanical
resonator.

Within the area of  superconducting nanomechanical systems, SQUID-based circuits have been intensively investigated.
A  dc SQUID  and its integration with the mechanical resonator has been theoretically proposed\cite{Zhou, Buks, Xue, Pugnetti,
Cohen} and experimentally demonstrated\cite{motiondetection,TunableBackaction,Etaki}. The detection of the mechanical
oscillations was possible through an analysis of the dynamics  of the electrical response of the superconducting circuit.
A SQUID is a non-linear cavity. Non-linear effects in opto- and nano-mechanical systems recently drew a lot of attention
(see Ref. \onlinecite{Aspelmeyer}). They originate from different sources --- non-linear coupling between the resonator and the
cavity, mechanical non-linearities of resonators made of carbon nanotubes and graphene, and also from the cavity itself.
The aim of this paper is to study the role of non-linearity in the dynamics of a SQUID coupled to a mechanical resonator. Our interest
is two-fold. On one hand, we would like to explore further the impact of mechanical oscillation to the stability of the SQUID dynamics,
thinking to eventually use this informalion as a detection mean. On the other hand,  backaction in  the presence of strong non-linearities
requires a closer inspection. Due to the complexity of dynamic behavior of non-linear systems one can expect that even small effect of
backaction can considerably alter the properties of a non-linear oscillator.

In this Article, we make the first step in this direction. We consider a dc SQUID coupled to a driven harmonic mechanical
resonator. Whereas formally this system is similar to the one studied experimentally in Ref. \onlinecite{motiondetection}, to explore
the regime where the backaction is the strongest, we consider the situation when the mechanical and the cavity frequencies are of
the same order. This is the regime when the dc SQUID acts as a Josephson parametric amplifier (JPA) \cite{Siddiqi,SQUIDamplifier}
and displays a multistable behavior: the amplitude of the oscillations of a driven JPA can assume two values in a wide frequency range,
such behavior was seen in the response of the non-linear oscillators to the parametric forcing~\cite{Lifshitz, Litak, parameticresonances}. This feature of the JPA can be used to readout the state of the qubit, since different amplitude correspond to different states of the qubit.
We consider the situation when both the field of the cavity (the phase of the Josephson junctions) and the coordinate of the mechanical
motion are classical variables. In our chosen setup, JPA is parametrically driven. Assuming a weak coupling between the JPA and the
mechanical oscillator, we demonstrate that the backaction can considerably modify the dynamics of JPA and lead to multistability
rather than bistability. We obtain this result analytically, approximating JPA by the Duffing oscillator, and also confirm it numerically,
relaxing this approximation. The results of this Article demonstrate that backaction can indeed essentially modify the behavior
of a non-linear oscillator. The present analysis is therefore relevant when considering the JPA as a detector for mechanical
motion. In comparison with the qubit detection, where the hysteretic regime permits a readout, the mechanical motion detection uses the whole finite amplitude regime. Therefore, here JPA acts as a threshold detector. Moreover, the multistability results in the extra hysteretic region, which enhances the sensitivity and range of the detector.  

The article is organized as follows. In Sec.~\ref{sec:system} we describe the device and set up the model to describe it.
In Sec. ~\ref{sec:EoM}, we consider the case of negligible  coupling (no backaction of the SQUID on the mechanical
oscillator) and solve the corresponding equations. In Sec.~\ref{sec:backaction}, the corrections to the amplitude due to
backaction are found by considering coupling term in the equations of motion. The conclusions of this work are summarized
in Sec.~\ref{sec:conclusion}.

\section{The Model}
\label{sec:system}

\begin{figure}
\centering
    \subfloat[]{\includegraphics[width=0.8 \linewidth]{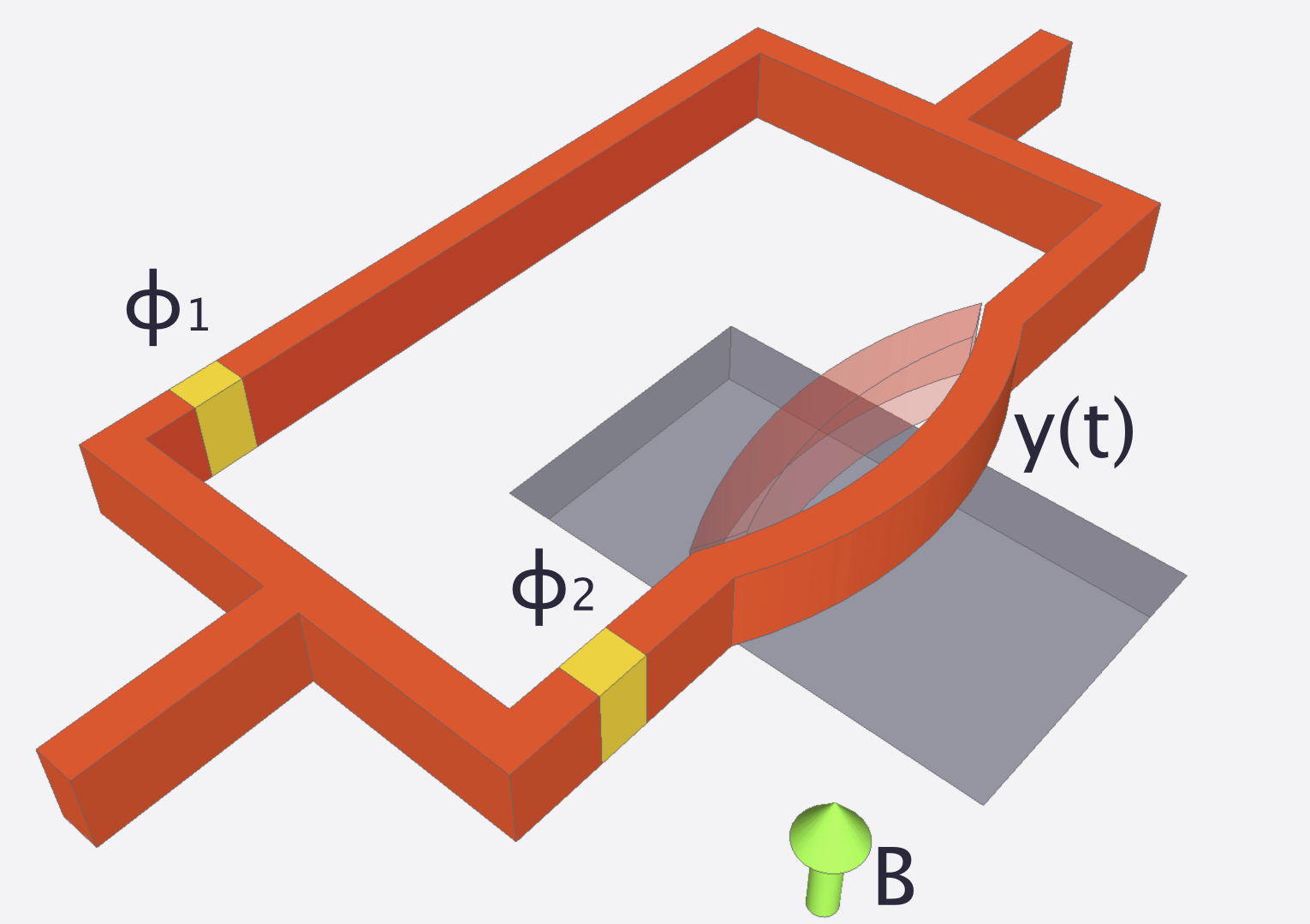}}\\
    \subfloat[]{\includegraphics[width=0.8 \linewidth]{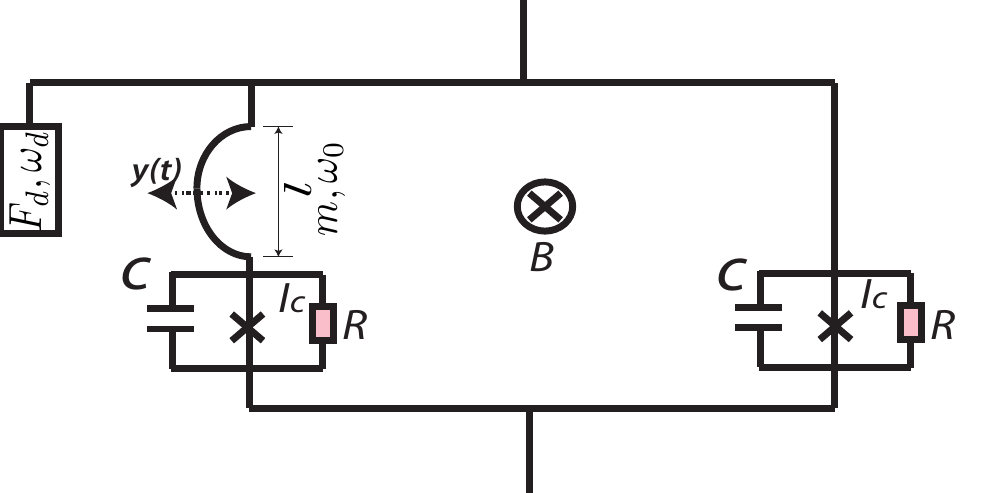}}
  \caption{(a) The picture of the system: the rectangular-shaped  dc SQUID with the suspended beam. The magnetic field is orthogonal to the SQUID. The oscillation of the beam is in the loop plane. Therefore, the displacement $y$ generates change in the magnetic flux passing through the loop. (b) The schematic overview of the setup. The system is driven with force $F_d$. Josephson junctions are modeled as resistively and capacitively shunted junctions(RCSJ).}
 \label{fig:SQUID}
\end{figure}

We consider a dc SQUID with two (nearly) identical Josephson junctions coupled to a mechanical resonator. The resonator is
formed by a suspended segment of superconductor. We only consider one mode of the mechanical resonator. It can be externally
driven, which is experimentally realized by fabricating the suspended part of the setup close to a piezoelectric element. The schematic overview of the system
is shown on  Fig.\ref{fig:SQUID}. The coupling between the SQUID and the mechanical resonator is based on the fact that the critical
current of the SQUID periodically depends on the magnetic flux, making it a very sensitive magnetic flux detector. The oscillations of
the resonator induce a variation of the area thus affecting the flux. When the mechanical resonator is driven, the flux modulation leads to a
parametric driving for the SQUID.\cite{Zhou, Buks}.

The mechanical resonator is modelled as a harmonic oscillator driven by external periodic force $F_d \cos(\omega_d t)$ with
driving frequency $\omega_d$. The magnetic flux passing through SQUID loop is dependent on the position of the
resonator, $\Phi=\Phi_{ext}+\beta_0 Bly$, where $\Phi_{ext}=\alpha_0 BA$ represents the flux through the area $A$ when resonator
is at rest,  $B$ is the magnetic field, and $y$ is the displacement from the equilibrium position of the resonator. The geometric
factors of $\alpha_0$ and $\beta_0$ (both of the order of unity) depend on the direction of the magnetic field. Finally the inductance of the
SQUID is assumed to be negligibly small.

To analyze the dynamics of this system one can look at the sum and difference of gauge-invariant phases across each Josephson
junction, respectively, $\varphi_\pm=(\phi_1\pm\phi_2)/2$. The condition that the superconductor order parameter is single
valued leads to the relation between phase difference and total flux bias
\begin{equation}
\varphi_-=\pi \frac{ \Phi}{\Phi_0}=\phi_e+\frac{y}{\xi}+2\pi n,
\label{phiminus}
\end{equation}
here $n$ is an integer, $\Phi_0 = \pi\hbar c/e$ is superconducting flux quantum, and the geometric constants of the system are
\begin{eqnarray}
\phi_e=\frac{\pi \alpha_0 BA}{\Phi_0} \hspace{10pt} \text{and}  \hspace{10pt}
\xi^{-1}=\frac{\pi \beta_0 Bl}{\Phi_0}.
\end{eqnarray}
The  Hamiltonian $H=H_y+H_\varphi+U$  of the SQUID with the movable arm is thus
\begin{align}
&H_y=\frac{p_y^2}{2m}+\frac{m\omega_0^2 y^2}{2}-F_d y \sin(\omega_d t),\\
&H_\varphi=\frac{p_{\varphi_+}^2}{2m_{\varphi_+}}, \label{Hvarphi}\\
& U=-2E_J\cos(\varphi_-)\cos(\varphi_+),
\end{align}
where $E_J=I_c \Phi_0/(2 \pi)$ is the Josephson energy, $p_y$, $m$ and $\omega_0$ are the momentum, the mass
and the frequency of the mechanical resonator, $p_{\varphi_+}$ and $m_{\varphi_+}$ are the momentum and the mass of the SQUID phase. The effective Hamiltonian is composed of the contribution from the
resonator  $H_y$, the kinetic energy of the SQUID $H_\varphi$, and the potential energy of the SQUID $U$. Since the
potential energy $U$ depends on the phase difference it provides the coupling between the SQUID and the oscillator through
Eq.(\ref{phiminus}). From our system it follows that the momentum of the mechanical oscillator is $p_y=m \dot{y}$. The phase momentum of the SQUID is given by $p_{\varphi_+}=m_{\varphi_+} \dot{\varphi}_+=E_J\dot{\varphi}_+^2/\omega_p$, where $m_{\varphi_+}=2E_J/\omega_p$ is the mass of the phase, $\omega_p=\sqrt{2 \pi I_c/(C \Phi_0)}$ is the plasma frequency,
$I_c$ and $C$ are the critical current and the capacitance of each Josephson junction.

Clearly, the potential energy of the SQUID is a non-linear function in $\varphi_-$ and $\varphi_+$.  Since we are interested in the two dynamical variables $y$ and $\varphi_+$, the dependence on the phase difference should be changed to the dependence on the oscillator displacement.
In the current experiments the ratio $y/\xi$ is very small which enables us to expand potential energy so that
\begin{equation}
U=-2E_J \left[ \cos(\phi_e)-\sin(\phi_e) \frac{y}{\xi} \right] \cos(\varphi_+).
\end{equation}

\section{Equations of motion}
\label{sec:EoM}

The dynamics of the system  is governed by the following equations of motion,
\begin{align}
&\ddot{y}+\lambda\dot{y}+\omega_0^2 y =\frac{F_d}{m} \cos(\omega_d t)
-\frac{2E_J \xi^{-1} \sin(\phi_e)}{m} \cos(\varphi_+),
\label{mechosc}\\
&\frac{\ddot{\varphi}_+}{\omega_p^2}+\frac{\delta}{\omega_p} \dot{\varphi}_++\left(\cos(\phi_e)-\sin(\phi_e)\frac{y}{\xi}\right)\sin(\varphi_+)=0.
\label{squid}
\end{align}
where $\lambda=\omega_0/ Q_y $ and $\delta=1/Q_\varphi$ are the dissipation rates for the resonator and the
SQUID, with $Q_y$ and $Q_\varphi$ being the respective quality factors.

The two equations are coupled by the last term in Eq. (\ref{mechosc}), which expresses the backaction of the
SQUID on the mechanical resonator, and by the term proportional to $y/\xi $ in Eq. (\ref{squid}), which provides
the effect of the resonator on the SQUID. Note that both couplings are proportional to $\xi^{-1}$. For $\xi^{-1}=0$, the
equations are decoupled: The SQUID is not driven, $\varphi_+ = 0$, whereas the driven mechanical resonator
shows the usual response,
\begin{equation}
\label{mechoscsol}
y(t)= A(\omega_d)\cos(\omega_d t)+ D(\omega_d)\sin(\omega_d t)
\end{equation}
with
\begin{eqnarray}
&&A(\omega_d)=\frac{F_d}{m}\frac{(\omega_0^2-\omega_d^2)}{(\omega_0^2-\omega_d^2)^2+\lambda^2\omega_d^2},\\
&&D(\omega_d)=\frac{F_d}{m}\frac{\lambda\omega_d}{(\omega_0^2-\omega_d^2)^2+\lambda^2\omega_d^2}.
\end{eqnarray}

In the following we will analyse the consequences of the coupling on the classical non-linear dynamics of the SQUID. We first
disregard the backaction of the SQUID (dropping the last term on the r.h.s  of Eq. (\ref{mechosc})) but still consider the effect of
the mechanical resonator on the SQUID. In this case, the SQUID is parametrically driven. There is always a
trivial (zero) solution for the overall phase-drop $\varphi_+$; the non-trivial one is found by substituting $y(t)$ from
Eq. (\ref{mechoscsol}). For convenience, we introduce the new parameters,
\begin{eqnarray}
&&\left\{\begin{array}{cc}
\omega=\frac{\omega_d}{2\omega_p}, \hspace{10 pt} \omega_{\phi}^2=\cos(\phi_e), \hspace{10 pt}\tau=\omega_p t,\\
\gamma=\xi^{-1}\sin(\phi_e)A(\omega), \hspace{10 pt}\beta=\xi^{-1}\sin(\phi_e)D(\omega),
\end{array}\right.
\end{eqnarray}
and choose the regime where $\cos(\phi_e)>0$, so that Eq.~\eqref{squid} is rewritten as
\begin{eqnarray}
\ddot{\varphi}_++\omega^2 \varphi_+ &&=-\delta \dot{\varphi}_++\omega^2\varphi_+ - \omega_{\phi}^2\sin(\varphi_+) \nonumber\\
&&+ \left[ \gamma\cos(2\omega \tau)+\beta\sin(2\omega \tau) \right] \sin(\varphi_+).
\end{eqnarray}
Under realistic experimental conditions , the coefficients in front of the phase dependent functions on the right-hand side
of equation are usually small. Setting them to zero gives the unforced and undamped linear oscillator.  Therefore,
we can use widely-used analytical techniques for solving Duffing oscillator~\cite{Duffingosc}, based on a perturbation
around a solution to the linear oscillator.  One of the methods to obtain the perturbative correction is based on the
van der Pol transformation,
\begin{eqnarray}
u=\varphi_+ \cos(\omega \tau)-\frac{\dot{\varphi}_+}{\omega}\sin(\omega \tau),\\
v=-\varphi_+ \sin(\omega \tau)-\frac{\dot{\varphi}_+}{\omega}\cos(\omega \tau),
\end{eqnarray}
where $u$ and $v$ are slowly varying quantities. The transformation turns the second-order differential equation into a
system of two first-order differential equations,
\begin{eqnarray}
\omega\dot{u}=&-&\sin(\omega \tau)\left[\omega\delta (u\sin(\omega \tau)+v\cos(\omega \tau))\right.\nonumber\\
&+&\omega^2\left(u\cos(\omega \tau)-v\sin(\omega \tau)\right)  \nonumber\\
&-& \omega_{\phi}^2\sin(\left(u\cos(\omega \tau)-v\sin(\omega \tau)\right))\nonumber\\
&+& (\gamma\cos(2\omega \tau) +\beta\sin(2\omega \tau))  \nonumber \\
        & &\times  \left.  \sin\left(u\cos(\omega \tau)-v\sin(\omega \tau)\right)\right],
\label{u}
\end{eqnarray}
\begin{eqnarray}
\omega\dot{v}= &-& \cos(\omega \tau)\left[\omega\delta (u\sin(\omega \tau)+v\cos(\omega \tau))\right.\nonumber\\
&+&\omega^2\left(u\cos(\omega \tau)-v\sin(\omega \tau)\right)  \nonumber\\
&- &\omega_{\phi}^2\sin(\left(u\cos(\omega \tau)-v\sin(\omega \tau)\right))\nonumber\\
&+&(\gamma\cos(2\omega \tau) +\beta\sin(2\omega \tau)) \nonumber\\
 & &\times  \left.  \sin\left(u\cos(\omega \tau)-v\sin(\omega \tau)\right)\right].
\label{v}
\end{eqnarray}
In order to solve such system the method of averaging over the period $T=2\pi/\omega$ is used. The idea is
to approximate equation in the form $\dot{x}=f(x,t)$ by averaging out fast oscillatory dynamics obtaining
$\dot{x'}= 1/T\int^T_0 f(x',t)dt\equiv \bar{f}(x')$. For that, slowly-varying quantities are written in polar
coordinates, i.e. $u=r\cos(\theta)$ and $v=r\sin(\theta)$. Then, the integrals of non-linear functions can
be found from the properties of the Bessel functions~\cite{besselfunctions},
\begin{eqnarray}
\int^{\pi}_{-\pi} e^{i (u \cos(\tau)+v \sin(\tau))} &&\cos(k \tau)d\tau\nonumber\\
&&=2\pi i^k\cos(k \theta) J_k(r),\\
\int^{\pi}_{-\pi} e^{i (u \cos(\tau)+v \sin(\tau))} &&\sin(k \tau)d\tau\nonumber\\
&&=2\pi i^k\sin(k \theta) J_k(r),
\end{eqnarray}
where $k$ is an integer and $J_k(r)$ is the Bessel function of the first kind. On proceeding this way we are left with two
equations in terms of the amplitude $r$ and the phase $\theta$,
\begin{eqnarray}
\omega \dot{r}&&=-\left(\delta \omega \frac{r}{2}+\frac{1}{2}(\gamma \sin(2\theta)+\beta\cos(2\theta))\left(J_1(r)+J_3(r)\right)\right),\nonumber\\
r \omega &&\dot{\theta}=-\left(-\omega_{\phi}^2J_1(r) +\omega^2 \frac{r}{2}\right.\nonumber\\
&&
\label{amplphasedifeq}
\qquad\left.+\frac{1}{2}(\gamma \cos(2\theta)-\beta\sin(2\theta))\left(J_1(r)-J_3(r)\right)\right).
\end{eqnarray}
The equilibrium points of the slow flow are determined by setting $\dot{r}$ and $\dot{\theta}$ to zero, giving
\begin{eqnarray}
& & \gamma \sin(2\theta)+\beta\cos(2\theta)=- \frac{\delta \omega r}{J_1(r)+J_3(r) } \;, \label{eqmd0a}\\
& & \gamma \cos(2\theta)-\beta\sin(2\theta)=-\frac{\omega^2 r-2 \omega_{\phi}^2 J_1(r)}{ J_1(r)-J_3(r)} \; .
\label{eqmd0b}
\end{eqnarray}
By taking the square and then summing Eqs. (\ref{eqmd0a}) and (\ref{eqmd0b}) one gets the algebraic equation for the
amplitude only,
\begin{equation}
\label{amplitude}
\frac{(r\omega'^2-2 J_1(r))^2}{\left( J_1(r)-J_3(r)\right)^2}+\frac{ \left(b r \omega'\right)^2}{\left( J_1(r)+J_3(r)\right)^2}=\frac{(\gamma ^2+\beta^2)}{4 \omega_{\phi}^4},
\end{equation}
where $\omega'=\omega/\omega_{\phi}$ and $b=\delta/\omega_{\phi}$.

We are interested in stable solutions of Eq. (\ref{amplitude}) and, therefore, we need to perform the stability analysis\cite{stability}. It is done by varying the amplitude and the phase of out-of-equilibrium solution $(r_0, \theta_0) $ obtained from Eqs. (\ref{eqmd0a}), (\ref{eqmd0b}),
\begin{equation}
\label{routofequil}
r=r_0+\Delta r \hspace{10 pt} \text{and} \hspace{10 pt} \theta=\theta_0+\Delta \theta,
\end{equation}
where $\Delta r$ and $\Delta \theta$ are small deviations. Substituting \eqref{routofequil} into \eqref{amplphasedifeq} and linearizing in $\Delta r$ and $\Delta \theta$, we obtain the constant coefficient system,
\begin{equation}
\left(\begin{array}{c}
\Delta \dot{ r}\\
\Delta \dot{\theta}
\end{array}\right)=\left(\begin{array}{cc}
e(r_0)&f(r_0)\\
 g(r_0)& h(r_0)

\end{array}\right)
\left(\begin{array}{c}
\Delta  r\\
\Delta \theta
\end{array}\right)={\textbf M} \left(\begin{array}{c}
\Delta  r\\
\Delta \theta
\end{array}\right) .
\end{equation}

We seek the following solutions to the linear system:  $\Delta r= A e^{\lambda t}$ and $\Delta \theta= D e^{\lambda t}$.
The solutions should be bound as $t$ goes to infinity, meaning the eigenvalues $\lambda$ of {\textbf M} should be negative.
We impose the conditions on  the trace of the associated matrix {\textbf M}, $tr=e(r_0)+h(r_0)\leq0$ and on its determinant,
$det=e(r_0)h(r_0)-f(r_0)g(r_0)>0$. From these conditions the stability testing of the solutions is straightforward.

\begin{figure}
\centering
\includegraphics[width=0.4\textwidth]{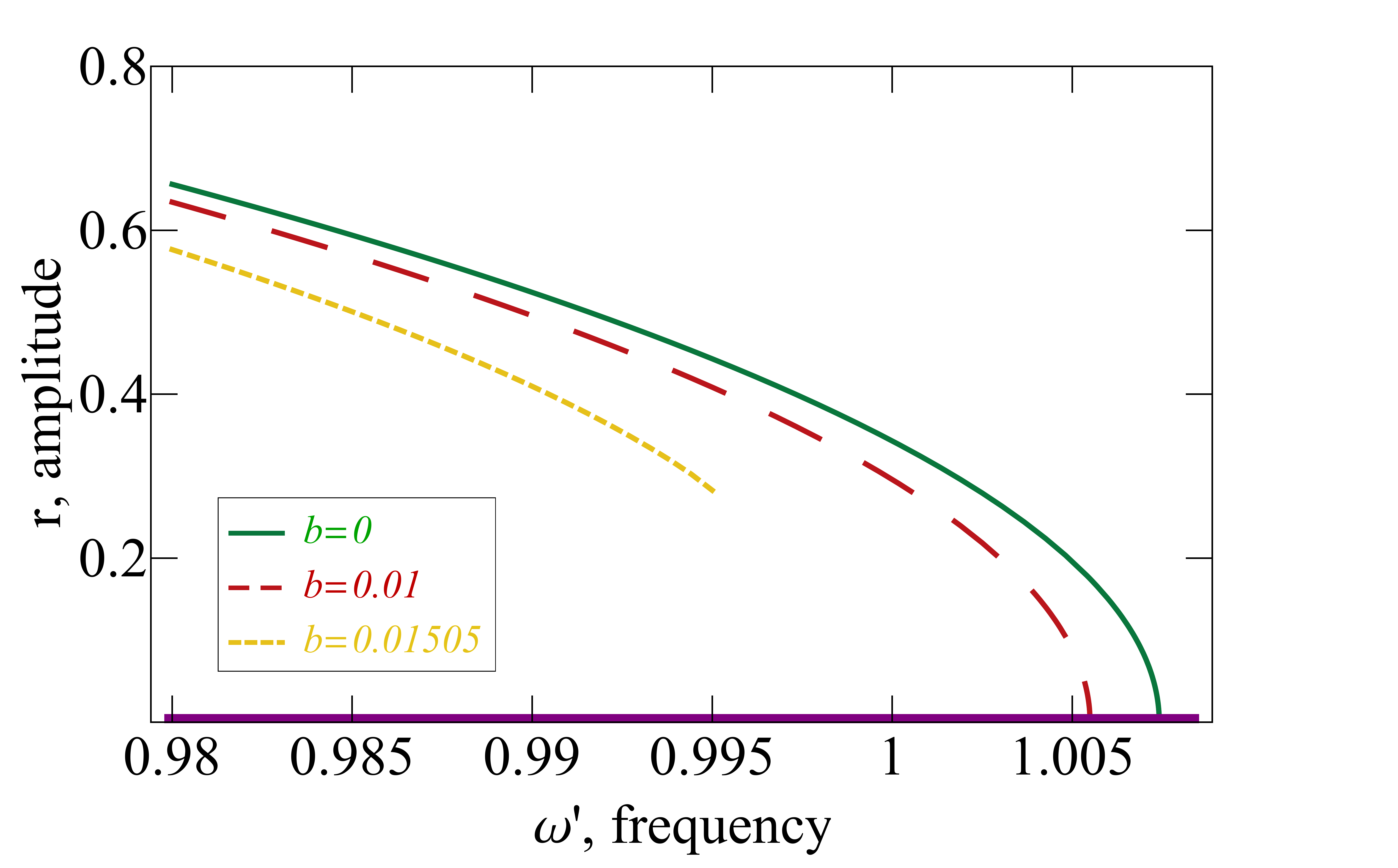}
\caption{(Color online) Frequency response of the SQUID for dimensionless driving force $c=0.03$ and different values of
SQUID dissipation $b$. Purple horizontal line represents zero solution. Here, only stable equilibria are shown.}
\label{fig:amplitude_nobackation}
\end{figure}

\begin{figure}
\centering
\includegraphics[width=0.4\textwidth]{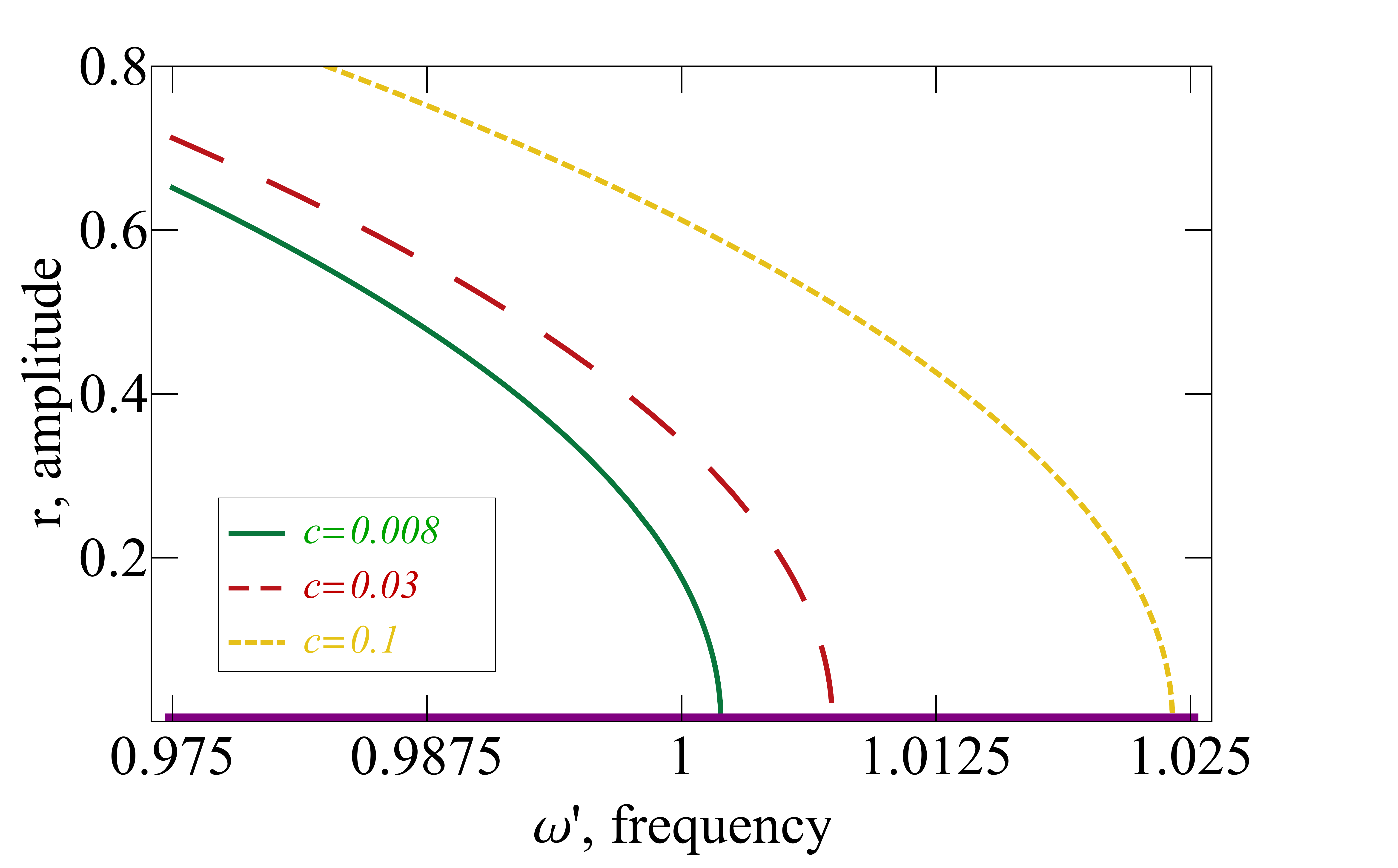}
\caption{(Color online) The dependence of the amplitude on the frequency for fixed dissipation $b=0.001$ and driving forces $c=0.008, 0.03, 0.1$ .}
\label{fig:amplitude_nobackation-2}
\end{figure}

Eq. (\ref{amplitude}) is the  equation for the amplitude $r$ if the backaction is not taken into account. The equation is
valid for any values of $r$. Analytical solutions can be obtained however only for small values of $r$. The denominators
on the left-hand side of Eq. \eqref{amplitude} can be safely ignored because the dissipation and quantity $\omega'^2-1$ are
small as long as the backaction is not included. Their inclusion does not change the result (they will be taken into account in the
next section for evaluating the effect of backaction). The Bessel function in the numerator of Eq. \eqref{amplitude} is instead
approximated up to the third order in the amplitude $J_1(r)\approx \frac{r}{2}-\frac{r^3}{16}$.
We focus on the  special case when mechanical oscillator is in the resonance $\omega_d\sim\omega_0$. Then,
introducing the dimensionless driving force $c=(\xi ^{-1}\tan(\phi_e) F_d)/(2m \omega_{\phi}  \omega_p\lambda)$ and
since $\gamma=0$, we obtain the amplitudes of the two stable solutions,
\begin{equation}
r_0= 2\sqrt{2-2\omega'^2+\frac{\sqrt{c^2-4b^2\omega'^4}}{\omega'}} \hspace{10 pt} \text{and} \hspace{10 pt} r_0=0.
\end{equation}
The amplitude  for different values of dissipation is shown in  Fig.~\ref{fig:amplitude_nobackation}. Note first that at
negligible dissipation and driving there is a bifurcation point $\omega_d = 2\omega_c$, which depends on the cavity
frequency   $\omega_c=\omega_p\sqrt{\cos(\phi_e)}$. Above this point, $r=0$ is the only stable solution; below this
point, a non-trivial solution emerges. This behavior is typical and has been experimentally observed for the parametrically
driven oscillator~\cite{parameticresonances}. The bistability below the bifurcation points must result in the hysteretic behavior.
If the frequency is slowly increasing the amplitude remains zero then "jumps up" to the upper branch follows it down.
If the frequency is sweeping backward, the amplitude follows the upper branch to some point and then drops to the lower branch.
To complete the picture we plot also the amplitude  for different values of the driving/coupling
(as parametrised by the parameter $c$). This is shown on  Fig.~\ref{fig:amplitude_nobackation-2}.

Both the finite dissipation and the finite driving push the bifurcation point to higher frequencies. However, when the dissipation
is too strong, $\omega'>\sqrt{c/(2 b)}$, the non-trivial solution does not appear. This condition is frequency dependent, and
therefore at the finite dissipation and weak driving the non-trivial solution does not exist close to the bifurcation point but
reappears at lower frequencies. Note that our analytical analysis is restricted to $r \ll 1$, therefore for strong enough
dissipation the non-trivial solution does not appear at any frequency.

\section{Backaction}
\label{sec:backaction}
\begin{figure*}
\centering
    \subfloat[]{\includegraphics[width=0.4 \linewidth]{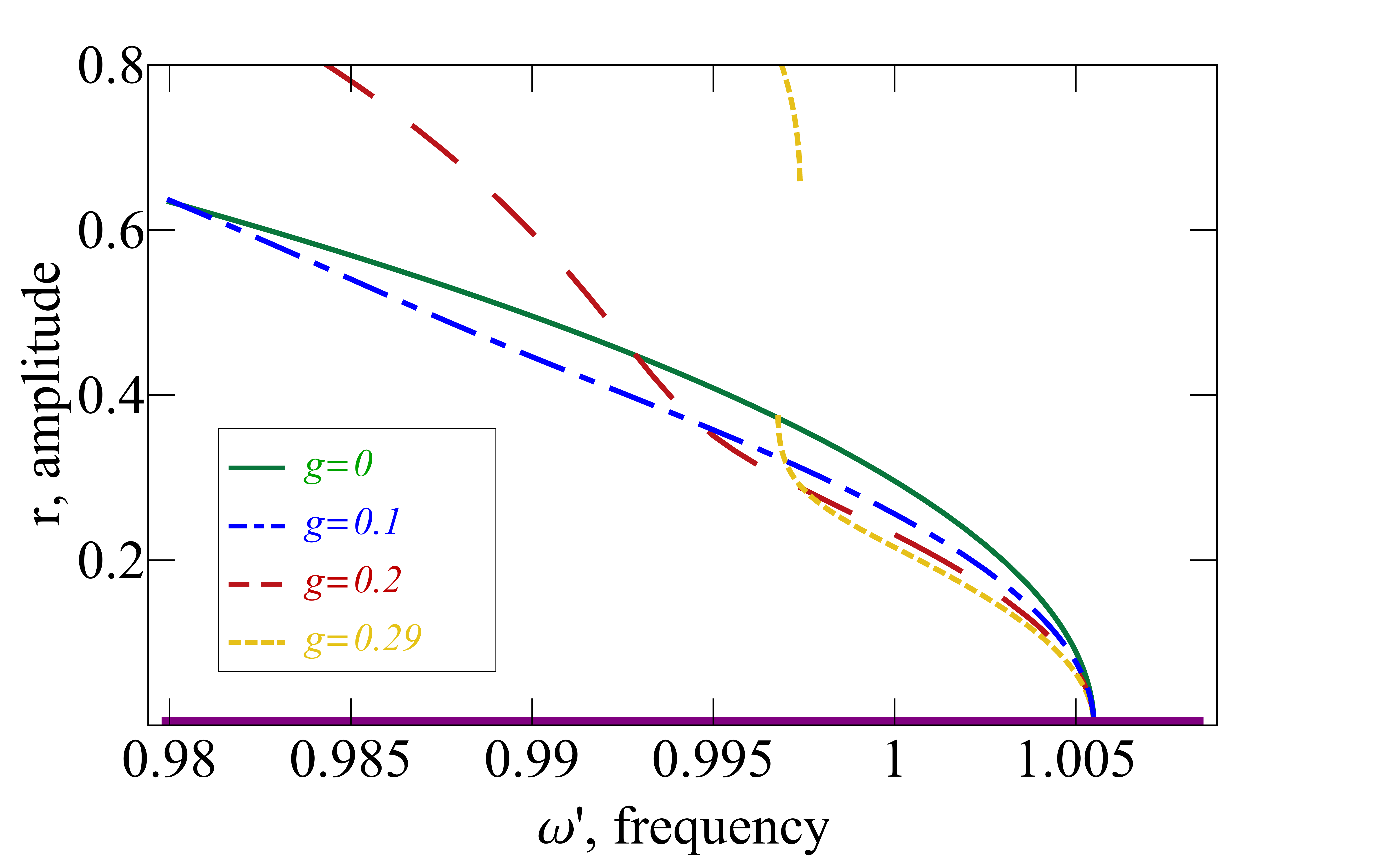}}
\quad
    \subfloat[]{\includegraphics[width=0.4 \linewidth]{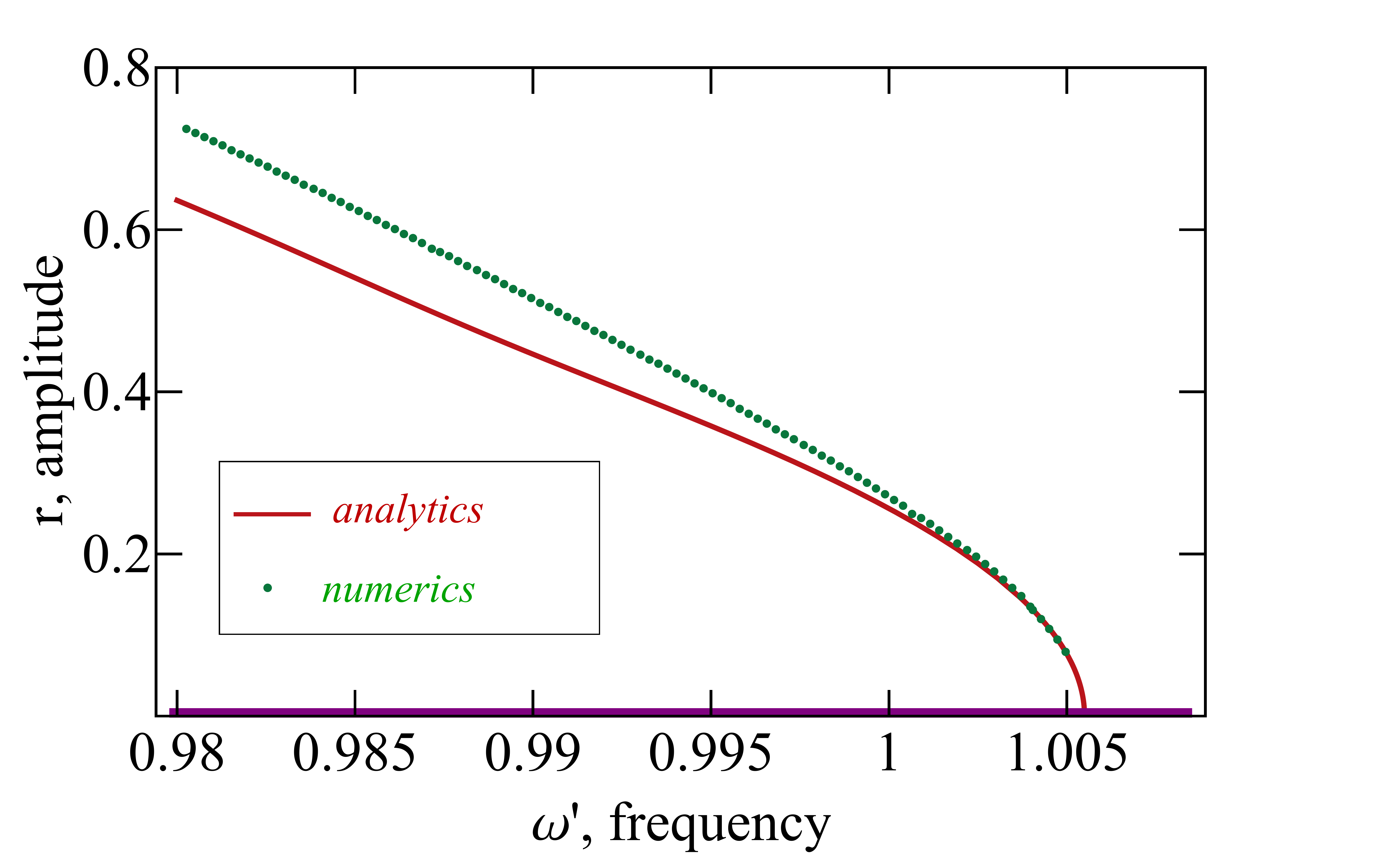}}
    \\
    \subfloat[]{\includegraphics[width=0.4 \linewidth]{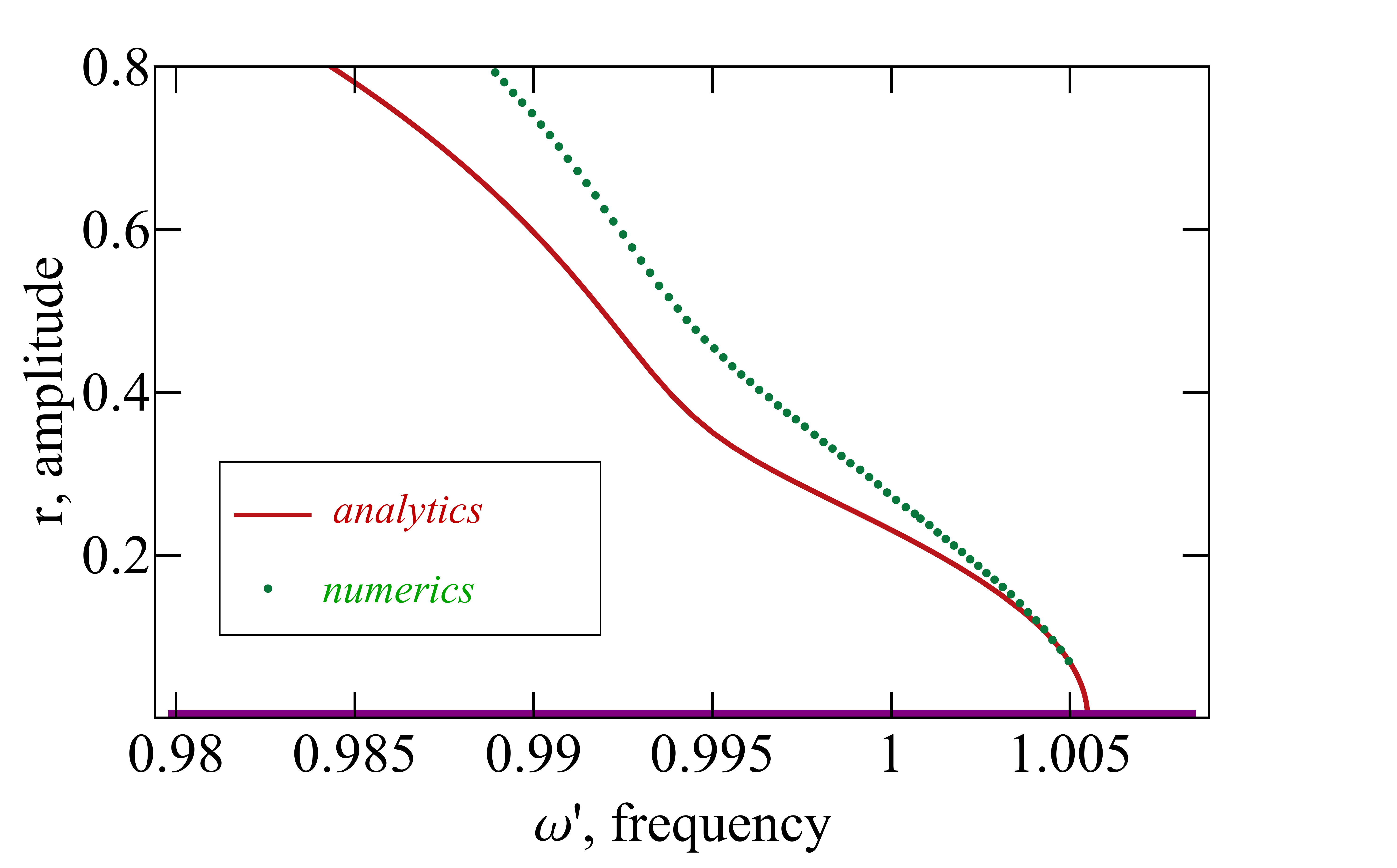}}
    \quad
    \subfloat[]{\includegraphics[width=0.4 \linewidth]{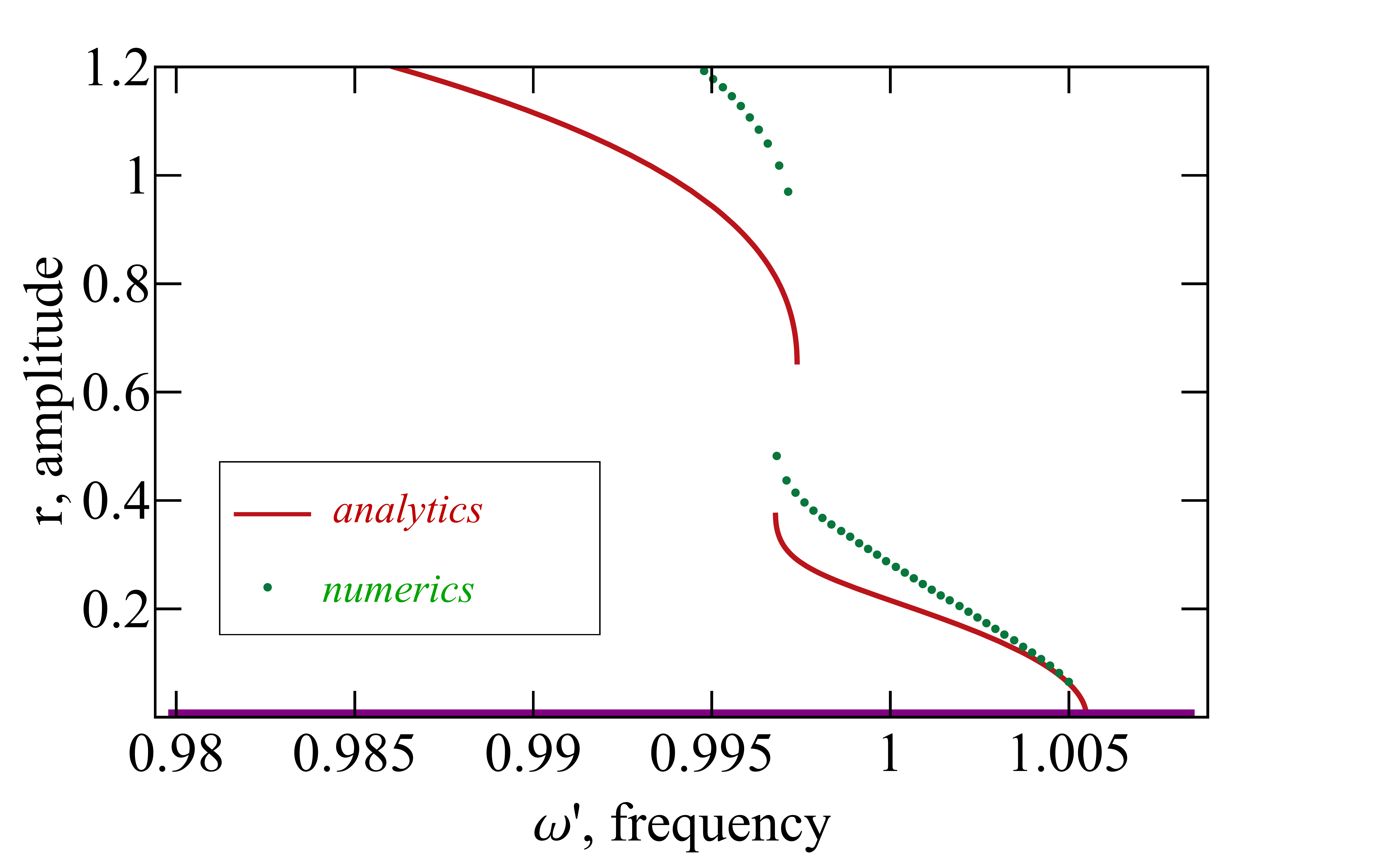}}
  \caption{(Color online)(a) The dimensionless amplitude $r$ of the dc SQUID overall phase is plotted {\it vs.} the frequency ratio $\omega'$ for different coupling strength $g$, which induces backaction. The values of dissipation is set to $b=0.01$ and the renormalized force is $c=0.03$. In (b)-(d), analytical amplitude results (solid line) are compared with numerical simulations (dotted line) for corresponding backaction strength (b) $g=0.1$, (c) $g=0.2$, and (d) $g=0.29$. Purple line shows the zero amplitude solution.}
  \label{fig:amplitude_backaction}
\end{figure*}

In Section \ref{sec:EoM}, we investigated the situation when backaction of the SQUID on the resonator is negligible. In this Section, we take this backaction perturbatively into account and demonstrate that it leads to further multistability of the SQUID. One can take into account the second term of Eq.~\eqref{mechosc} by inserting solutions from Section \ref{sec:EoM}, $\varphi_+=r \cos(\theta)\cos(\omega_d t/2)-r\sin(\theta)\sin(\omega_d t/2)$. In this way, we perturbatively study the effect of backaction on the amplitude of the SQUID.

We can eliminate the phase dependence using  Eqs.~\eqref{eqmd0a} by $\cos(2\theta)=-b {\omega'}^2 r/c(J_1+J_3)$ and then expanding $\cos(\varphi_+)$ in term of the Bessel functions using the identities
\begin{eqnarray}
&&\cos\left(p \cos(\frac{\omega_d t}{2})\right)\cos\left(m\sin(\frac{\omega_d t}{2})\right)=\nonumber\\
&&2\sum^\infty_{i=0}(-1)^i \left[J_{2i}\left(p \right) J_{2i+2}\left(m \right)-J_{2i+2}\left(p \right) J_{2i}\left(m \right)\right]\cos(\omega_d t)\nonumber
\end{eqnarray}

and
\begin{eqnarray}
&&\sin\left(p \cos(\frac{\omega_d t}{2})\right)\sin\left(m\sin(\frac{\omega_d t}{2})\right)=\nonumber\\
&& 2\sum^\infty_{i=0}(-1)^i\left[J_{2i+1}(p) J_{2i+3}(m)-J_{2i+3}(p) J_{2i+1}(m)\right ]\sin(\omega_d t)\nonumber\\
\label{coscos_sinsin}
&&\qquad\qquad\qquad\qquad\qquad\qquad +2 J_1(p)J_1(m) \sin(\omega_d t),
\end{eqnarray}
where $p=r\cos(\theta)$ and $m=r\sin(\theta)$. Other harmonics of the overall phase which have the frequencies $ \omega=n \omega_d/(2\omega_p)$, where $n$ is integer, are disregarded in the expansions.  They generate the Bessel functions of higher orders, which are small for the amplitudes of our interest $r \ll 1$.

To provide an analytical solution, we again look at small amplitudes. In this case the two cosines of eq. \eqref{coscos_sinsin} can be approximated by $(m^2-p^2)\cos(\omega_d t)/4$, and the two sines give $pm\sin(\omega_d t)/2$. Then, we see that the driving force is shifted due to the backaction
\begin{equation}
F_d \rightarrow F'=F_d-\frac{E_J  \sin(\phi_e)}{c \xi}\frac{b {\omega'}^2 }{1-\frac{r^2}{12}} r^2 \ ,
\end{equation}
and the additional sinusoidal force is generated
\begin{equation}
G=-\frac{E_J \sin(\phi_e)r^2}{c \xi }\sqrt{\frac{1}{4} c^2-\frac{b^2 {\omega'}^4}{(1-\frac{r^2}{12})^2}}.
\end{equation}
Hence, the equation for the resonator becomes
\begin{equation}
\ddot{y}+\lambda \dot{y}+\omega_0^2 y=\frac{F'}{m}\cos(\omega_d t)+\frac{G}{m}\sin(\omega_d t).
\end{equation}
This equation has the same solutions as driven harmonic oscillator given in Eq.~\eqref{mechoscsol} with the modified amplitudes,
\begin{equation}
A \rightarrow A'=-\frac{G }{ m\lambda\omega_d}\hspace{5pt}\text{and} \hspace{5pt} D\rightarrow D'=\frac{F' }{ m\lambda\omega_d}.
\end{equation}

\noindent
The shift in the  oscillator's amplitudes affects the quantities $\gamma$ and $\beta$ in the equation of motion for the SQUID. It follows that the backaction is included for the amplitude of the SQUID through the quantity
\begin{align}
\frac{\gamma^2+\beta^2}{\omega_{\phi}^4}=\left(\frac{ \xi^{-1} \tan(\phi_e)}{2m \lambda \omega_{\phi} \omega_p \omega'}\right)^2\left(F'^2+G^2\right) \ .
\end{align}
The equation for the amplitude becomes
\begin{equation}
\label{ampbackaction}
\frac{(\omega'^2-1 +\frac{r^2}{8})^2}{\left(\frac{1}{2}-\frac{r^2}{12}\right)^2}+\frac{ \left(b \omega'\right)^2}{\left(\frac{1}{2}-\frac{r^2}{24}\right)^2}=\frac{c^2+ g^2 r^4}{\omega'^2}- \frac{4 g b r^2}{1-\frac{r^2}{12}}
\end{equation}
where $g=E_J \xi^{-2} \tan(\phi_e) \sin(\phi_e)/(4 m \omega_{\phi} \omega_p \lambda)$ is the strength of backaction. Note that this is still an algebraic equation though more sophisticated than the expansion of Eq. (\ref{amplitude}).

Fig. ~\ref{fig:amplitude_backaction} (a) displays the stable non-trivial solution for the phase oscillation amplitude $r$ to Eq.~\eqref{ampbackaction} as a function of the frequency $\omega'$ for different values $g$ of backaction and for the finite dissipation $b=0.01$. One can see that there is still a bifurcation point, and its position is not affected by the backaction. The effect of the backaction is to suppress the amplitude close to the bifurcation point. However, further from this point the amplitude $r$ enhances. As the backaction gets stronger it  changes the monotonous behavior of the amplitude. In the certain range of frequencies the non-trivial solution becomes multi-stable: at the same frequency, there might be two stable non-trivial solutions in addition to the trivial solution $r=0$.

Note, however, that our analytical treatment, based on the quartic expansion of the cosine potential (the Duffing model), is only valid for $r \ll 1$. The non-trivial features that we observe appear at $r \sim 0.5$ and are strictly speaking outside the range of applicability of our approximation. To check whether they really exist we perform the numerical analysis of the full Hamiltonian of the system, still assuming weak backaction ($g \ll 1$), but without expanding the cosine. We solve numerically the system of equations \eqref{mechoscsol} and \eqref{squid}. The solution is valid for any values of the amplitude $r$. The results are presented in
Fig.~\ref{fig:amplitude_backaction} (b)-(d). We take the initial conditions for solving the differential equations to be our analytical solutions. The time evolved from 0 to 20000 and after it reached the steady-state we measured the amplitude. For small enough amplitudes the analytical results coincide with numerical ones. For higher amplitudes there is certain deviation, however the pattern stays the same. In particular, the numerical solution confirms the multistability.

\subsection{Josephson parametric amplifier as a coordinate detector}
\label{precision}

\begin{figure*}
\centering
    \subfloat[]{\includegraphics[width=0.32 \linewidth]{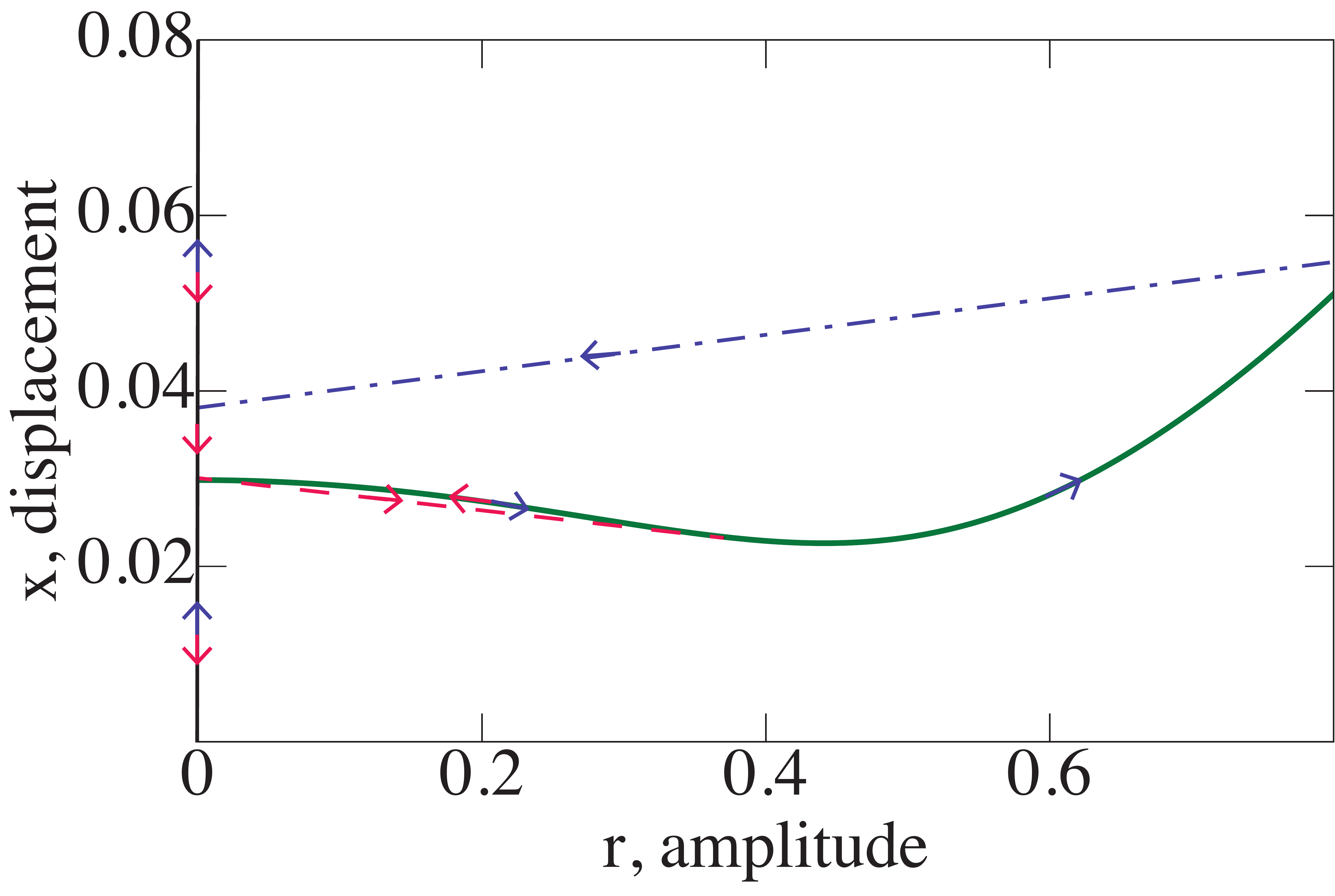}}
\quad
    \subfloat[]{\includegraphics[width=0.32 \linewidth]{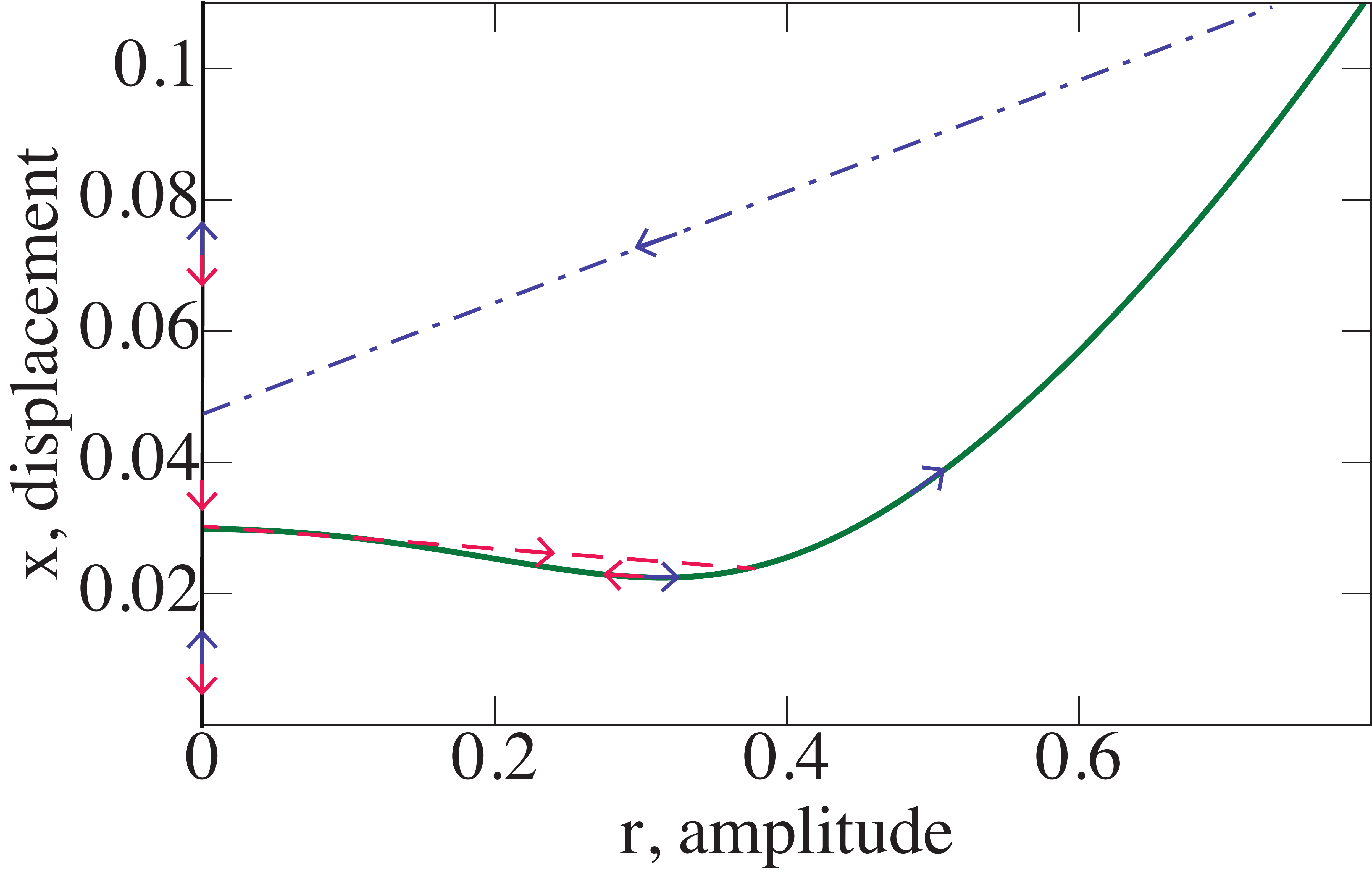}}
\quad
    \subfloat[]{\includegraphics[width=0.32 \linewidth]{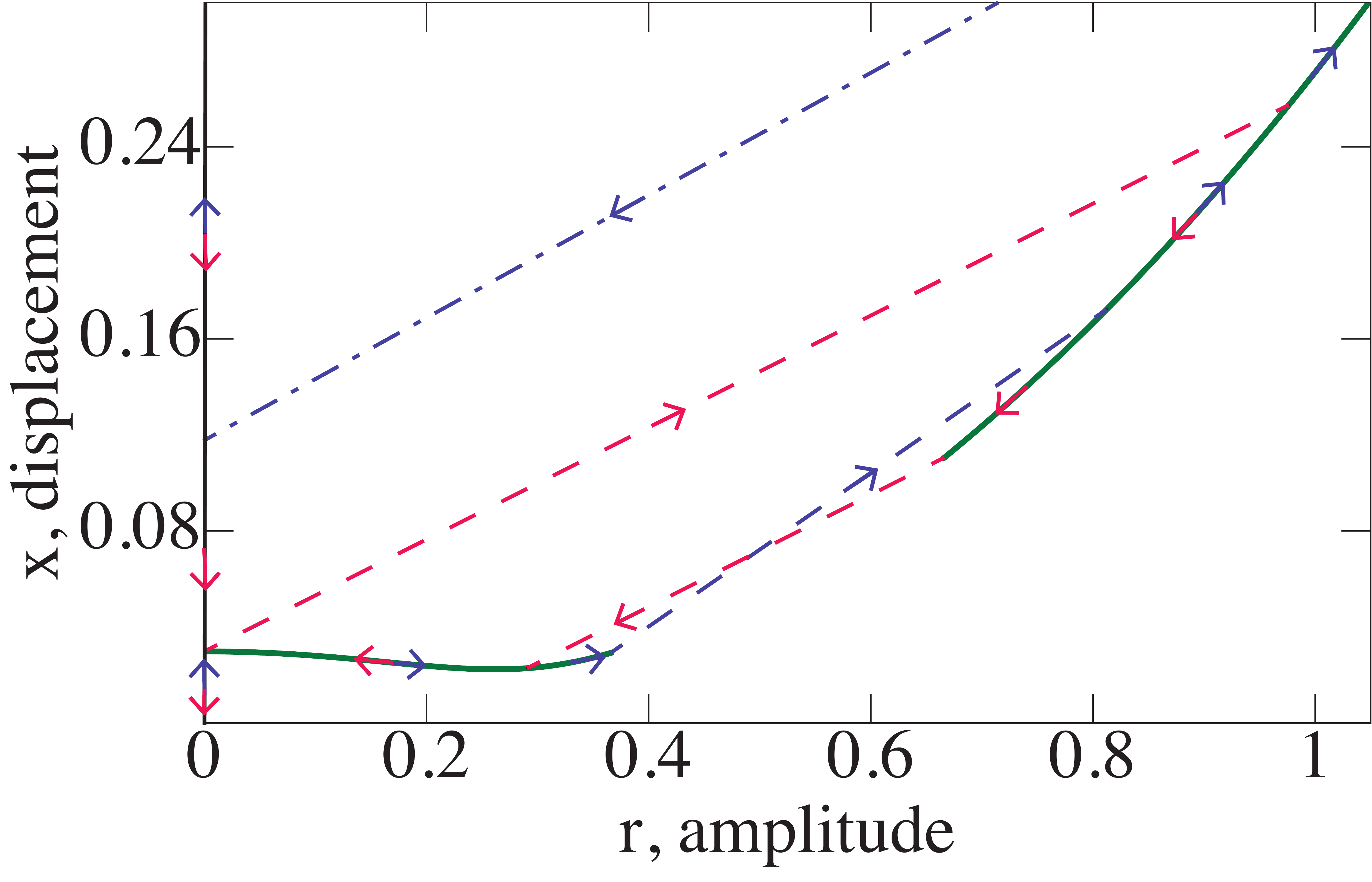}}

  \caption{(Color online) The dimensionless mechanical displacement $x$ as a function of the amplitude $r$ of the dc SQUID for different coupling strength (a) $g=0.1$, (b) $g=0.2$, and (c) $g=0.29$.  The arrows show the direction for the forward (red) and backward (blue) frequency swapping.  The dashed lines represent the amplitude jumps between two stable solutions. The dashed-dotted line is approximate location for the high amplitude bistable point.}
  \label{fig:displacement_amplitude}
\end{figure*}

One of the most popular applications of Josephson parametric amplifier is the readout of the qubit state \cite{Siddiqi}. This application uses the fact that when JPA is in a bistable regime, different values of the amplitude may be coupled differently to two states of the qubit, and thus a projective measurement of the amplitude of JPA yields the information of the occupied qubit state. In this subsection, we analyze how JPA can be used for the readout of the amplitude of the oscillations of the mechanical resonator it is coupled to. Whereas we are only considering a classical problem here, the next step would be to investigate the oscillator in the quantum regime, when the amplitude becomes quantized, and this is why we have chosen the oscillation amplitude as the parameter of interest.

For this purpose, we redrew our results as the dependence of the dc SQUID amplitude on the resonator oscillation amplitude. The results are shown in Fig.~\ref{fig:displacement_amplitude} for three different values of the coupling, where the dimensionless displacement $x=\xi^{-1}\sin(\phi_e) \sqrt{A'^2+D'^2}$ is plotted as a function of $r$. The arrows show a picture of the system evolvement between metastable states(zero and finite amplitude) for each frequency swapping. The exact location of the high amplitude bistable point is outside the scope of our article and can be determined by means of singularity analysis. Therefore, we plotted it as an approximated dashed-dotted line.

Fig. ~\ref{fig:displacement_amplitude}a which corresponds to weak backaction, shows that JPA can only detect the amplitude in a (narrow) range of values, where the curve is not flat. In contrast to the qubit detection, here it is best to use not the hysteretic range, which is narrow, but the whole branch corresponding to the non-zero amplitude $r$. Thus, JPA serves as a threshold detector. If one increases the coupling, the curve becomes steeper, and the range where detection is possible, increases. Fig. ~\ref{fig:displacement_amplitude} c, which corresponds to even stronger coupling and has an additional hysteretic region, shows that in this case, the detection is possible in two separate regions, separated by the bifurcation region. The sensitivity of the detector is thus enhanced by the appearance of the additional bifurcation. 

\section{Conclusions}
\label{sec:conclusion}

In this Article, we considered a SQUID coupled to a driven linear mechanical resonator. We found that above the bifurcation point, the phase of the SQUID does not respond to the parametric driving. The bifurcation point is located at the frequency slightly above the point where the condition $\omega_d = 2\omega_c$ is met. The precise location of the bifurcation point depends on the dissipation and the driving force, but not on the backaction strength. Below the bifurcation point, a non-trivial solution for the amplitude of the SQUID oscillation arises. We found that the dissipation suppresses this non-trivial solution whereas the driving strength enhances it. Furthermore, we discover that at strong enough backaction, the non-trivial solution may become multistable, and thus the physics of the systems goes beyond the Duffing oscillator. We have obtained the multistability by analytical calculations within the Duffing oscillator approximation and also by numerical solution of the full model. We demonstrated that this system can be used as the displacement detector by translating SQUID amplitude into mechanical displacement. Multistability increases the range and the sensitivity of the detector. 

One can experimentally control the strength of backaction by changing the geometry of the resonator or parameters of the SQUID. In our model the magnetic field was applied perpendicular to the SQUID loop, instead one can use parallel orientation of magnetic field and tune flux $\Phi_{ext}$ using {\em e.g.} a stripline, which also changes the backaction strength.

Currently, the most stringent condition in our theory which hinders the experimental verification is that the frequencies of the SQUID and the mechanical resonator are of the same order. In the existing experiments~\cite{motiondetection,TunableBackaction,Etaki}, the plasma frequency of the SQUID was several orders of magnitude higher than the mechanical frequency. However, currently there is an interest to the fabrication of superconducting junctions with suspended carbon nanotubes~\cite{Schneider} and graphene sheet integrated into the microwave cavities \cite{Vibhor}. Whereas the frequency of mechanical motion in the existing devices is still lower than the cavity frequency, the regime $\omega_d \sim 2\omega_c$ can be achieved. Another limitation is that we assumed the two Josephson junctions to be identical. We do not expect however that the asymmetry of the SQUID would qualitatively affect our results.

Finally, we emphasize that we have only made the first step towards exploring non-linear cavity properties in cavity electrodynamics. We certainly expect more rich and interesting physics in the situations which lies outside the scope of our Article --- strong backaction (beyond the perturbation theory), dispersive coupling between the SQUID and the resonator, as well as quantum effects in both the phase of the SQUID and the mechanical motion of the resonator.

\section*{Acknowledgments}

This work was supported by the Netherlands Foundation for Fundamental Research on
Matter (FOM), by Italian MIUR via PRIN Project 2010LLKJBX and by EU through IP-SIQS.

\end{document}